\documentclass[a4paper,11pt]{article}

\usepackage{pos}
\usepackage{bm}
\usepackage[utf8]{inputenc}

\title{Electromagnetic and axial-vector structure of singly heavy
  baryons in a pion mean-field approach} 

\author*[a]{Hyun-Chul Kim}

\affiliation[a]{Department of Physics, Inha University, \\
    Incheon 22212, Republic of Korea}

\emailAdd{hchkim@inha.ac.kr}

\abstract{In this talk, we present a series of recent works on the
  electromagnetic and axial-vector structures of low-lying singly
  heavy baryons. We first explain the pion mean-field approach, in
  which light and singly heavy baryons can be considered on an equal
  footing. We then discuss the results for the electromagnetic and
  radiative transition form factors of the singly heavy baryons. We
  also demonstrate the results for the strong decay rates and quark
  spin content of the singly heavy baryons. Finally, we propose a
  consistent way of dealing with the $1/m_Q$ corrections in the pion
  mean-field approach.}

\FullConference{The XVIth Quark Confinement and the Hadron Spectrum
  Conference (QCHSC24)\\  19-24 August, 2024
 Cairns Convention Centre, Cairns, Queensland, Australia
}

\begin{document}
\maketitle
\section{Introduction}
\begin{figure}[htp]
  \centering
   \includegraphics[scale=0.9]{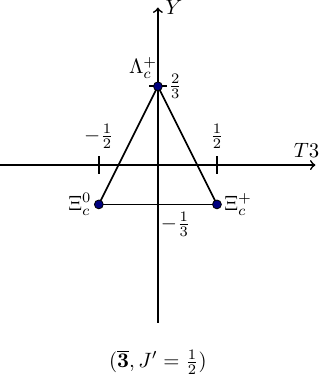}
  \includegraphics[scale=0.9]{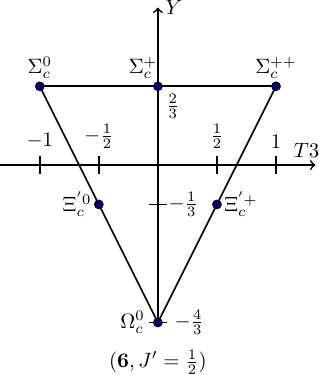} 
  \includegraphics[scale=0.9]{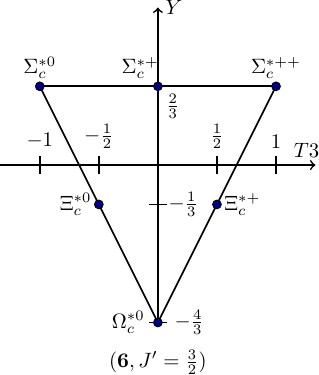}
  \caption{SU(3) representation of lowest-lying singly heavy baryons}
  \label{fig:1}
\end{figure}

A singly heavy baryon contains two light quarks and one heavy
quark. If we take the limit of the infinitely heavy quark mass, i.e.,
$m_Q\to \infty$, the heavy-quark spin, $\bm{S}_Q$ can not be flipped,
which leads to the conservation of $\bm{S}_Q$. So, the spin of the
light quarks is also conserved: $\bm{S}_L\equiv \bm{S}-\bm{S}_Q$. This
is called ``heavy-quark spin symmetry''~\cite{Isgur:1989vq,
  Isgur:1991wq, Georgi:1990um}, which makes the total spin of the
light quarks a good quantum number. Moreover, in the limit of $m_Q\to
\infty$, the heavy quark is independent of flavor, which results in
``heavy-quark flavor symmetry.'' Thus, the heavy quark inside a singly
heavy baryon plays a mere role of a static color source. It only
contributes to the total spin of the singly heavy baryon. This
indicates that dynamics inside the singly heavy baryon is governed by
the light quark degrees of freedom. Consequently, the low-lying singly
heavy baryons can be classified by flavor
$\mathrm{SU(3)}_{\mathrm{f}}$ symmetry: $\bm{3}\otimes \bm{3} =
\bar{\bm{3}}\oplus \bm{6}$. The total spin of the light quarks is
given by $\bm{\frac12}\otimes \bm{\frac12} = \bm{0}\oplus
\bm{1}$. When it is coupled to heavy quark, one can get the total
spins of the singly heavy baryons as $J'=1/2$ and $J'=3/2$. The baryon
antitriplet ($\bar{\bm{3}}$) and sextet are depicted in the 
left and right panels of Fig.~\ref{fig:1}, respectively. 
It is natural for the baryon antitriplet to have $J'=1/2$, and for the 
baryon sextet to have either $J'=1/2$ and $J'=3/2$. The baryon
sextet are degenerate. The degeneracy can only be lifted by
considering the $1/m_Q$ corrections. We will later discuss how the baryon
antitriplet has $J'=1/2$ whereas the baryon sextet has either $J'=1/2$
and $J'=3/2$ within the chiral quark-soliton model~($\chi$QSM). 

Witten proposed a seminal idea that an ordinary light baryon can be
viewed as the bound state of $N_c$ valence quarks in a meson mean
field, if one considers the large $N_c$ (the numober of colors)
limit. The presence of the $N_c$ valence quarks will polasize the
vacuum, and generate the pion mean field. Then, the $N_c$ valence
quarks are bound by the pion mean field in a self-consistent
manner. The $\chi$QSM can be considered as a realization of this
Witten's idea~\cite{Diakonov:1987ty, Christov:1995vm, Diakonov:1997sj}.  
The very same idea can be applied to the singly heavy baryon. 
If we strip off one light quark and replace it with a heavy quark,
then the singly heavy baryon can be regarded as a state of the
$N_c-1$ valence quarks bound by the pion mean field, where have been
produced by the presence of the $N_c-1$ valence
quarks~\cite{Yang:2016qdz, Kim:2018xlc, Kim:2019rcx}. Thus, the pion
mean-field approach or the $\chi$QSM allows one to describe both the
light and singly heavy baryons in an equal footing (see also a
review~\cite{Kim:2018cxv}). 

The mass spectrum of the singly heavy baryons was successfully
reproduced~\cite{Yang:2016qdz, Kim:2018xlc, Kim:2019rcx}, compared
with the experimental data. We found that the pion mean field for the
singly heavy baryons becomes weaker than that for the light
baryons. Since the baryon sextet are degenerate, we introduced the
$1/m_Q$ chromomagnetic hyperfine interactions to remove the
degeneracy. Using the same theoretical framework, we extended to
compute various properties of the singly heavy baryons. In this talk,
we summarize a series of recent results for the electromagnetic (EM) and
axial-vector observables and discuss them.
\section{Pion mean-field approach}
\begin{figure}[htp]
  \centering
  \includegraphics[scale=0.15]{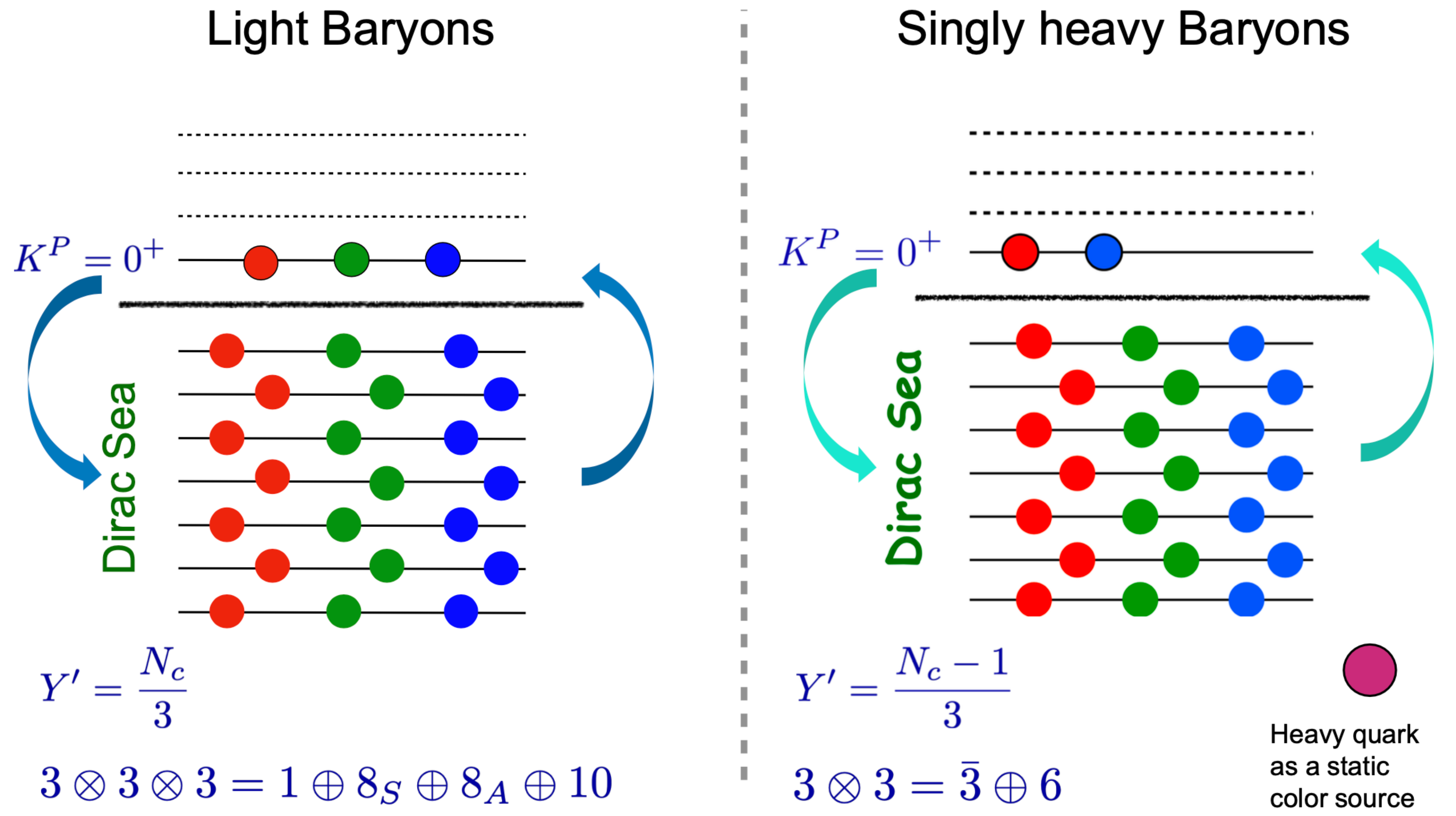}
  \caption{Pion mean-field picture: Light baryons and singly heavy baryons}
  \label{fig:2}
\end{figure}
Figure~\ref{fig:2} depicts schematically the mean-field picture of
both light and singly heavy baryons. The pion mean field for the
singly heavy baryons is created by the presence of the $N_c-1$ valence
quarks, whereas the heavy quark remains as a static color
source. Thus, the low-lying singly heavy baryons can be clasified as
representations of flavor $\mathrm{SU}_f(3)$ symmetry. 
For example, a singly heavy baryon $\Sigma_c^+$ can be identified as
the state with $J=1/2$, $T=1$, $T_3=0$, and $Y=2/3$. The right
hypercharge $Y'$ for singly heavy baryons, which will later be
explained in detail, is constrained by the number of the valence
quarks. As shown in Fig.~\ref{fig:2}, the right hypercharges for the
light and heavy baryons are given as $Y'=N_c/3$ and $Y'=(N_c-1)/3$,
respectively. They allow one to have the lowest-lying
representations: the baryon octet ($\bm{8}$) and decuplet ($\bm{10}$)
for the light baryons, and the baryon antitriplet ($\bar{\bm{3}}$),
sextet ($\bm{6}$) and so on~\cite{Yang:2016qdz, Kim:2018cxv}. 

The $\chi$QSM starts with the effective chiral action 
defined by  
\begin{align}
S_{\mathrm{eff}} := -N_c \mathrm{Tr} \ln \left[ i\rlap{/}{\partial}  +
  iMU^{\gamma_5} + i   \hat{m}\right], 
\end{align} 
where $\mathrm{Tr}$ consists of the functional trace over
four-dimanional space-time, and the traces over flavor and spin
spaces. $M$ is the dynamical quark mass that arises from the
spontaneous breakdown of chiral symmetry. 
The $U^{\gamma_5}$ represents the chiral field defined by
\begin{align}
  \label{eq:1}
U^{\gamma_5} (z) = \frac{1-\gamma_5}{2}   U(z) + U^\dagger(z)
  \frac{1+\gamma_5}{2} 
\end{align}
with
\begin{align}
  \label{eq:2}
U(z) = \exp[{i\pi^a(z) \lambda^a}],  
\end{align}
where $\pi^a(z)$ stands for the pseudo-Nambu-Goldstone (pNG) fields
and $\lambda^a$ are the flavor Gall-Mann matrices. 
$\hat{m}$ designates the mass matrix of current quarks $\hat{m} =
\mathrm{diag}(m_{\mathrm{u}},\, m_{\mathrm{d}},\,m_{\mathrm{s}})$.
Note that we deal with the strange current quark mass $m_{\mathrm{s}}$
perturbatively. Thus, we will consider it when we make a zero-mode
quantization for a collective baryon state.  

The one-particle Dirac Hamiltonian is given as 
\begin{align}
  H = \gamma_4 \gamma_i \partial_i + \gamma_4 MU^{\gamma^5}
    + \gamma_4 \overline{m} \mathbf{1},
\end{align}
where $\overline{m}$ is the average value of the current up and down
quark masses. Solving the following eigenvalue equation 
\begin{align}
H \psi_n(\bm{x}) = E_n \psi_n(\bm{x}),  
\end{align}
we obtain the energy eigenvalues of the single-quark state, $E_n$, and
the corresponding eigenfunctions $\psi_n(\bm{x})$. Then, we can
compute the baryon correlation 
function $\langle J_B (y)\Psi_h(y) (-i\Psi^{\dagger}_h(x)\gamma_4)
J_B^\dagger(x)\rangle_0$ as follows~\cite{Kim:2019rcx}:  
\begin{align}
\langle J_B (y)\Psi_h(y) (-i\Psi_h^\dagger(x) \gamma_4)
J_B^\dagger(x)\rangle_0 &  \sim \exp\left[-\{(N_c-1)E_{\mathrm{val}} +
  E_{\mathrm{sea}}+m_Q\}T\right] = \exp[-M_B T],  
\label{eq:6}
\end{align}
where the classical mass of a singly heavy baryon is obtained to be 
\begin{align}
M_B = (N_c-1) E_{\mathrm{val}} + E_{\mathrm{sea}} + m_Q.
\end{align}

To assign the quantum numbers to the classical baryon, we need to
quantize them. This can be done by considering the zero-mode
quantization, which will preserve the mean-field
solution~\cite{Christov:1995vm, Yang:2016qdz}.  
Having carried out the quantization, we arrive at the collective
Hamiltonian for singly heavy baryons 
\begin{align}
H =& H_{\mathrm{sym}} + H^{(1)}_{\mathrm{sb}},
\label{eq:Hamiltonian}
\end{align}
where $H_{\mathrm{sym}}$ is the flavor SU(3) symmetric part
\begin{align} 
H_{\mathrm{sym}}=M_{\mathrm{cl}}+\frac{1}{2I_{1}}\sum_{i=1}^{3}
\hat{J}^{2}_{i} +\frac{1}{2I_{2}}\sum_{a=4}^{7}\hat{J}^{2}_{a}.  
\label{eq:sym}
\end{align} 
Here, $I_{1}$ and $I_{2}$ are the moments of inertia of the
soliton. The explicit expressions for $I_{1,\,2}$ are given in
Appendix A in Ref.~\cite{Kim:2018xlc}.
The operators $\hat{J}_{i}$ and $\hat{J}_{a}$ are the spin generators in
SU(3). In the $(p,\,q)$ representation of the SU(3) group, we obtain the
eigenvalue of the SU(3) quadratic Casimir operator $\sum_{i=1}^8 
J_i^2$ as    
\begin{align}
C_2(p,\,q) = \frac13 \left[p^2 +q^2 + pq + 3(p+q)\right].   
\end{align}
Thus, the eigenvalues of $H_{\mathrm{sym}}$ are obtained as  
\begin{align} 
E_{\mathrm{sym}}(p,q) = M_{\mathrm{cl}}+ \frac{1}{2I_{1}} J(J+1) 
+\frac{1}{2I_{2}}\left[C_2(p,\,q) - J(J+1)\right] 
-\frac{3}{8I_{2}} Y'^2.
\label{eq:RotEn}
\end{align} 
The right hypercharge $Y'$ is constrained to be $(N_c-1)/3$,
which is imposed by the $N_c-1$ valence quarks inside a singly heavy
baryon. In the Skyrme model the right hypercharge is
constrained by the Wess-Zumino term. The right hypercharge is given by 
$Y'=2/3$, so that allowed representations are the baryon antitriplet
($\overline{\bm{3}}$), sextet ($\bm{6}$) with $J=1/2$ and $J=3/2$,
antidecapentaplet ($\overline{\bm{15}}$) with $J=1/2$ and $J=3/2$,
and so on. 

The pion mean-field solution must be coupled to the heavy quark. 
Thus, the collective wave functions of the baryons are obtained to be 
\begin{align} 
\psi_B^{({\mathcal{R}})}(J'J'_3,J;A)=\sum_{m_s=\pm1/2} C_{J_Qm_3
  JJ_3}^{J'J_3'} \sqrt{\mathrm{dim}(p,\,q)} (-1)^{-\frac{ \overline{Y}  }{2}+J_3}
  D^{(\mathcal{R})\ast}_{(Y,T,T_3)(\overline{Y}  ,J,-J_3)}(A),  
\label{eq:12}
\end{align} 
where 
\begin{align}
\mathrm{dim}(p,\,q) = (p+1)(q+1)\left(1+\frac{p+q}{2}\right).  
\end{align}
$J$ and ${J_{3}}$ are the spin and its third component of the light
quark degress of freedom, respectively. 

The symmetry-breaking part of the collective
Hamiltonian is given by
\begin{align} 
H^{(1)}_{\mathrm{sb}} 
=&\frac{\Sigma_{\pi N}}{m_{0}}\frac{m_{\mathrm{s}}}{3}
+\alpha D^{(8)}_{88}+ \beta \hat{Y}
+ \frac{\gamma}{\sqrt{3}}\sum_{i=1}^{3}D^{(8)}_{8i}
\hat{J}_{i},
\label{eq:sb}
\end{align}
where
\begin{align} 
\alpha=\left (-\frac{\Sigma_{\pi N}}{3m_0}+\frac{
  K_{2}}{I_{2}} Y  
\right )m_{\mathrm{s}},
 \;\;\;  \beta=-\frac{ K_{2}}{I_{2}}m_{\mathrm{s}}, 
\;\;\;  \gamma=2\left ( \frac{K_{1}}{I_{1}}-\frac{K_{2}}{I_{2}} 
 \right ) m_{\mathrm{s}}.
\label{eq:alphaetc}
\end{align}
We refer to Ref.~\cite{Kim:2018xlc} for the explicit expressions for
the moments of inertia and the $\pi N$ sigma term.

\section{Electromagnetic observables of singly heavy baryons}
The EM current is written as 
\begin{align}
  \label{eq:16}
J_\mu (x) = \bar{\psi} (x) \gamma_\mu \hat{\mathcal{Q}} \psi(x) + e_{Q}
  \bar{\Psi}   \gamma_\mu \Psi,  
\end{align}
where $\hat{\mathcal{Q}}$ denotes the charge operator in 
$\mathrm{SU_{f}(3)}$, defined by
\begin{align}
 \label{eq:17}
\hat{\mathcal{Q}} =
  \begin{pmatrix}
   \frac23 & 0 & 0 \\ 0 & -\frac13 & 0 \\ 0 & 0 & -\frac13
  \end{pmatrix} = \frac12\left(\lambda_3 + \frac1{\sqrt{3}}
                                                  \lambda_8\right). 
\end{align}
Here, $\lambda_3$ and $\lambda_8$ are the flavor SU(3) Gell-Mann
matrices. The $e_Q$ in the second part of the EM current
in Eq.~(\ref{eq:16}) denotes the heavy-quark charge, which
is $e_c=2/3$ for the charm quark or $e_b=-1/3$ for the
bottom quark.  Since the magnetic form factor of a heavy quark is
proportional to the inverse of the corresponding heavy-quark mass,
i.e., $\bm{\mu} \sim (e_Q/m_Q) \bm{\sigma}$, the heavy-quark term can
be ignored in the limit of $m_Q\to \infty$. Note that a heavy quark
inside a singly heavy baryon only gives a constant contribution to its
electric form factor in the $m_Q\to \infty$ limit.

\begin{figure}[htp]
  \centering
  \includegraphics[scale=0.2]{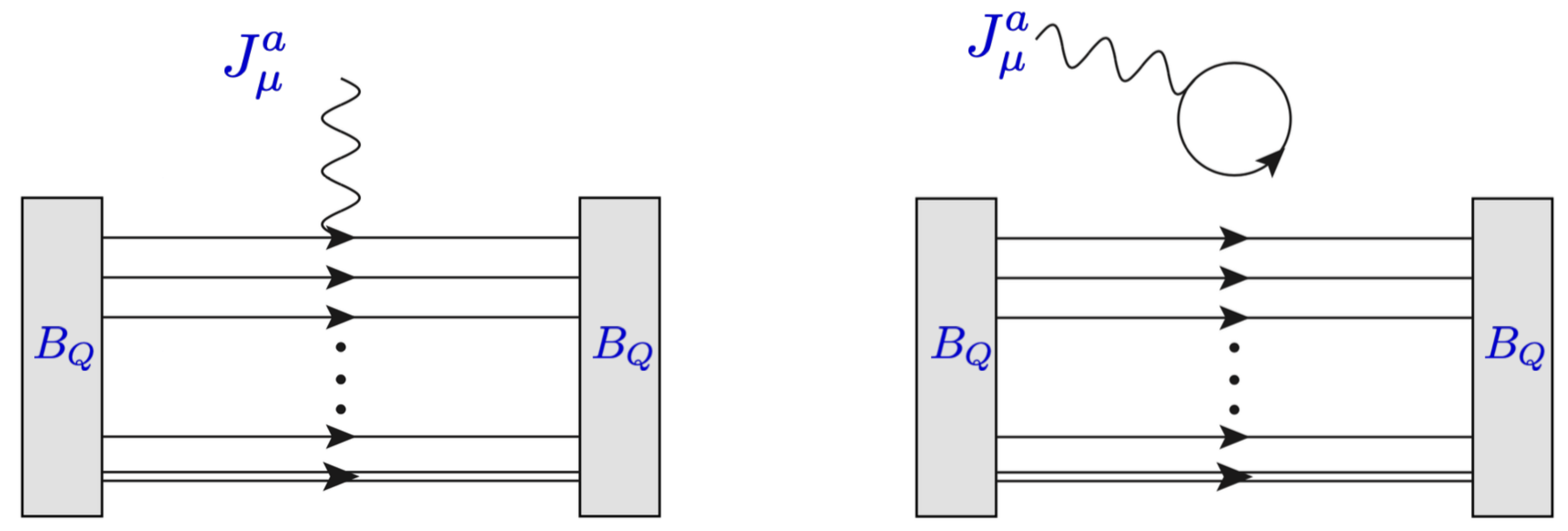}
  \caption{Three-point correlation function in the $\chi$QSM. }
  \label{fig:3}
\end{figure}
To compute EM and axial-vector observables of baryons, we
have to compute the three-point correlation functions, which are
schematically shown in Fig.~\ref{fig:3}. For a detailed formalism, we
refer to Ref.~\cite{Kim:1995mr,Kim:2018nqf}. 

\begin{figure}[htp]
  \centering
  \includegraphics[scale=0.2]{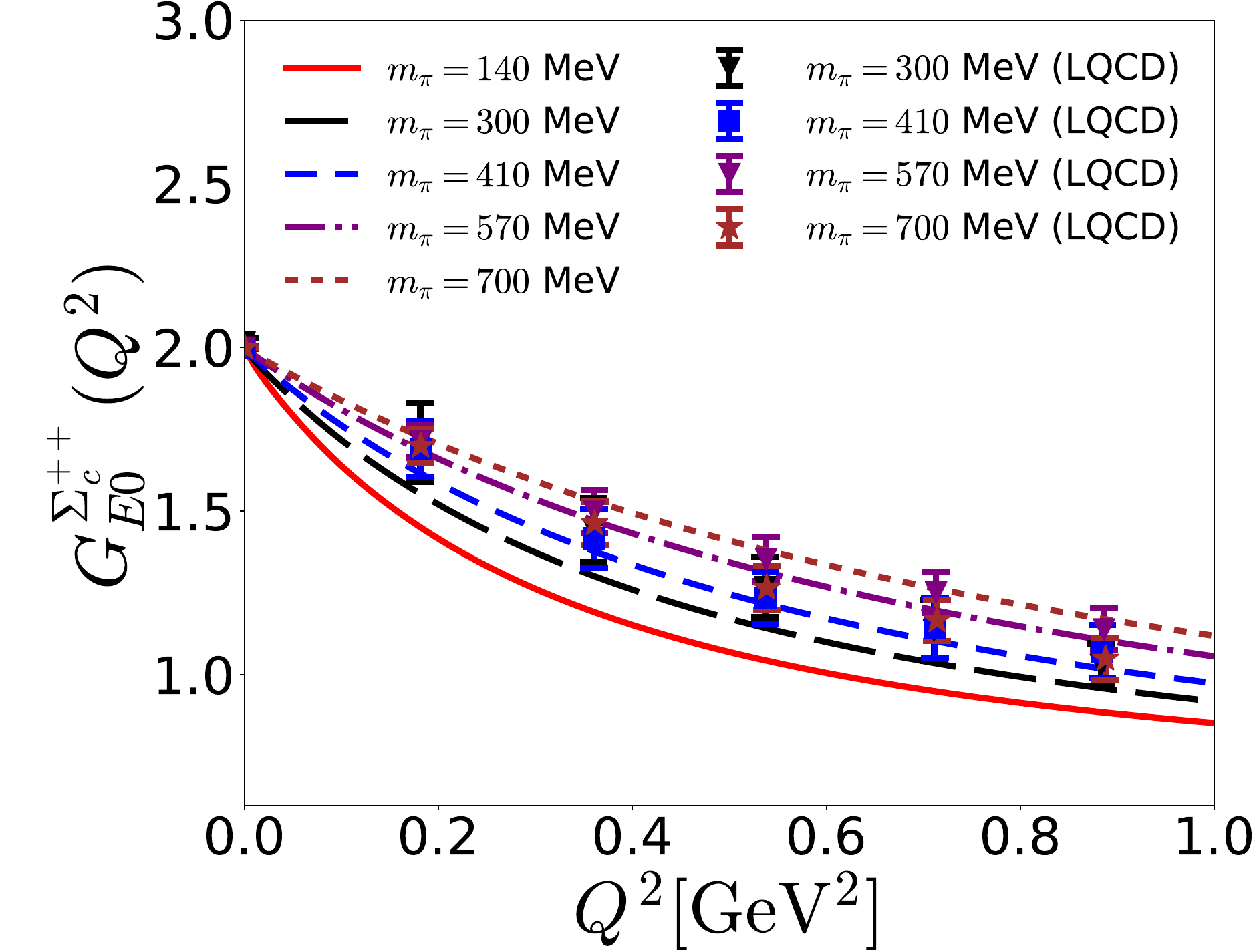}
  \includegraphics[scale=0.2]{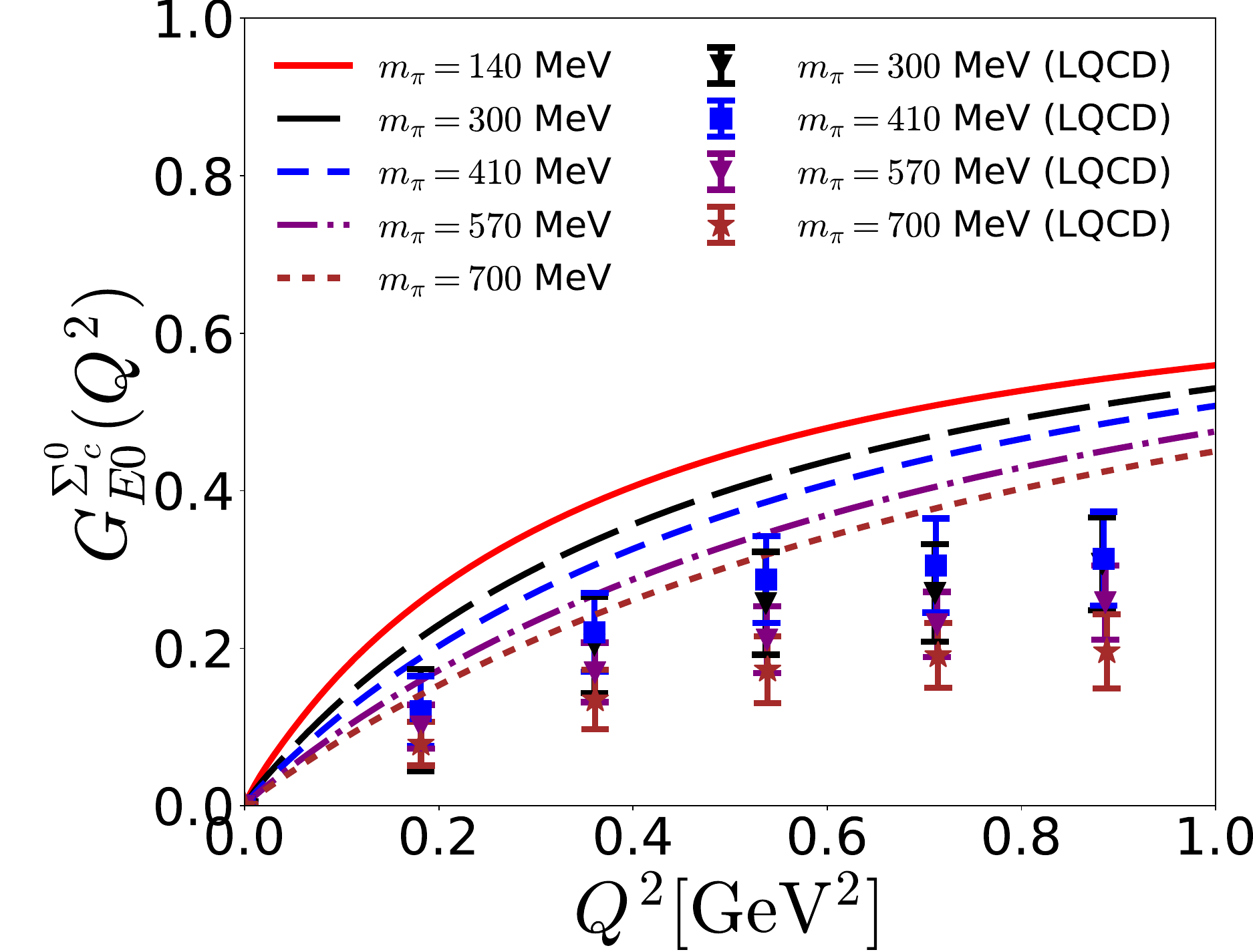}
  \includegraphics[scale=0.2]{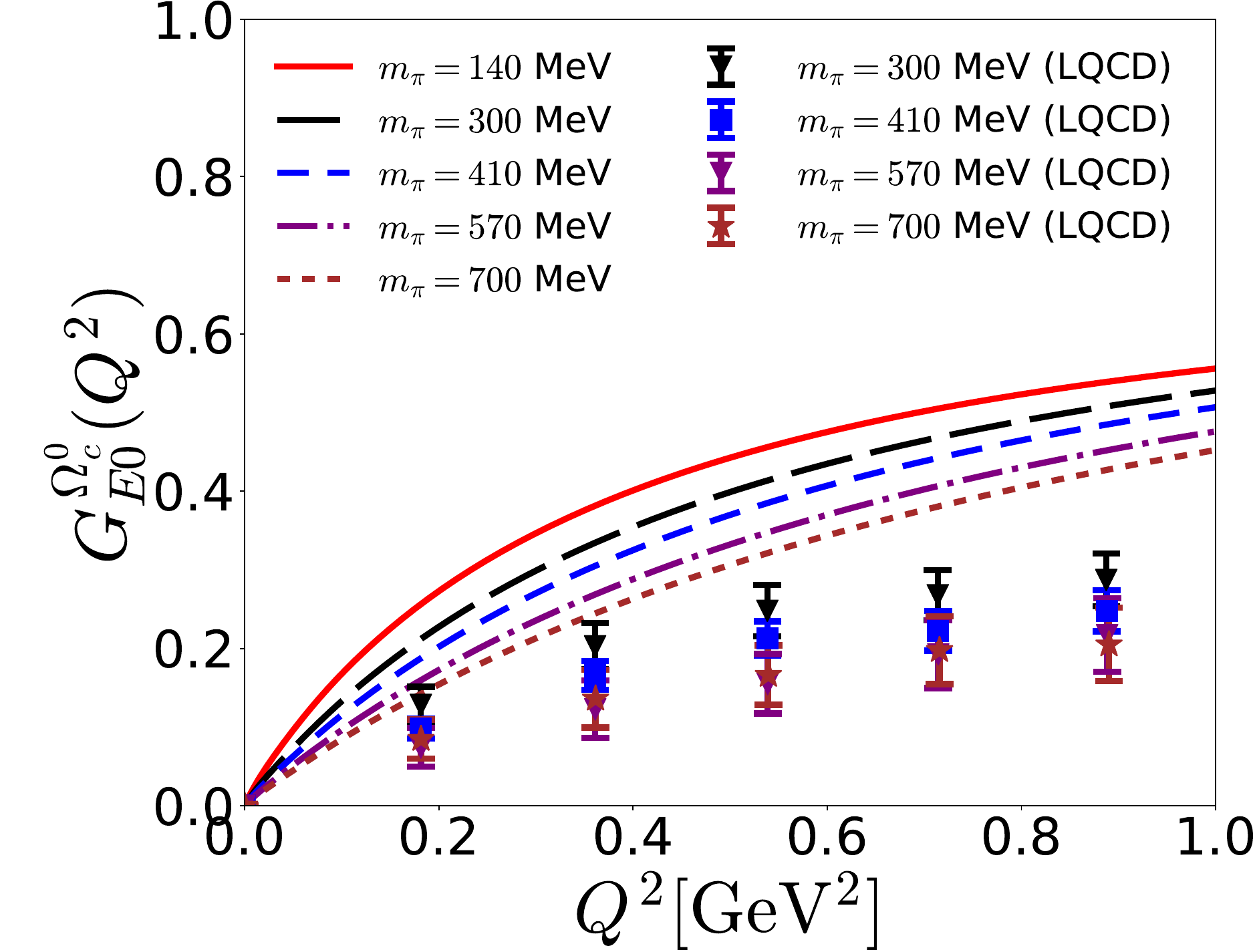}
  \caption{Results for the electric form factors of the singly heavy baryons.}
  \label{fig:4}
\end{figure}

\begin{figure}[htp]
  \centering
  \includegraphics[scale=0.2]{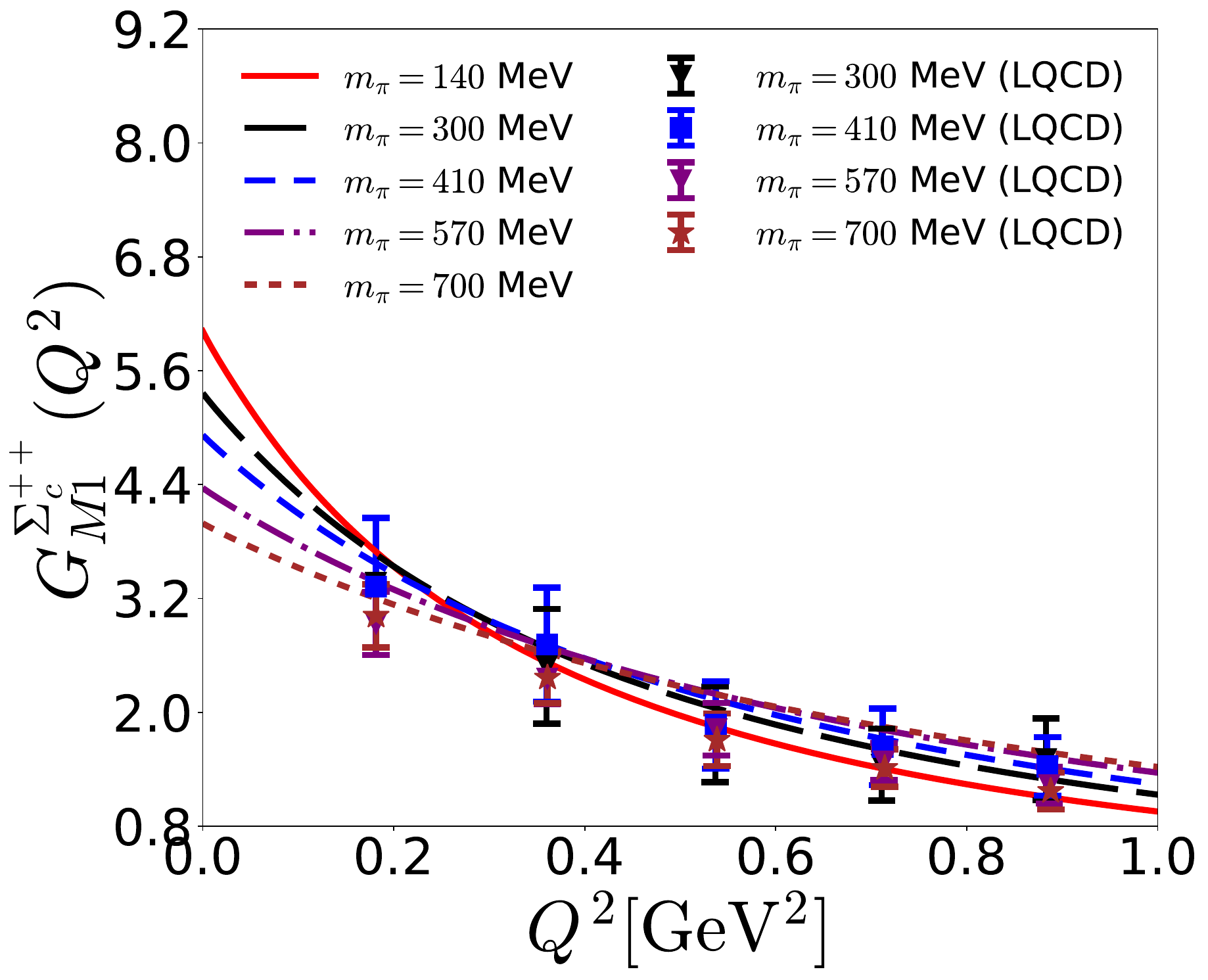}
  \includegraphics[scale=0.2]{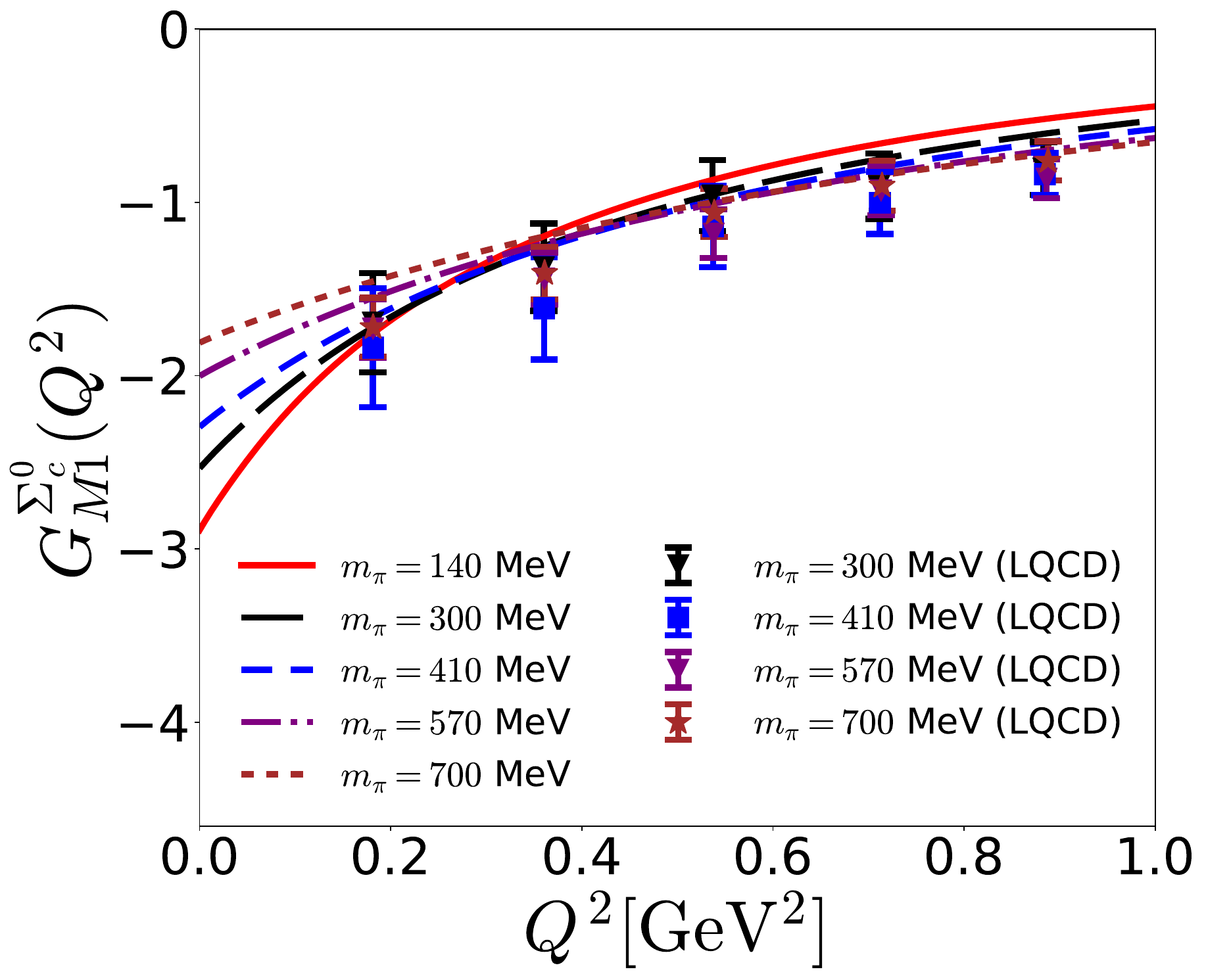}
  \includegraphics[scale=0.2]{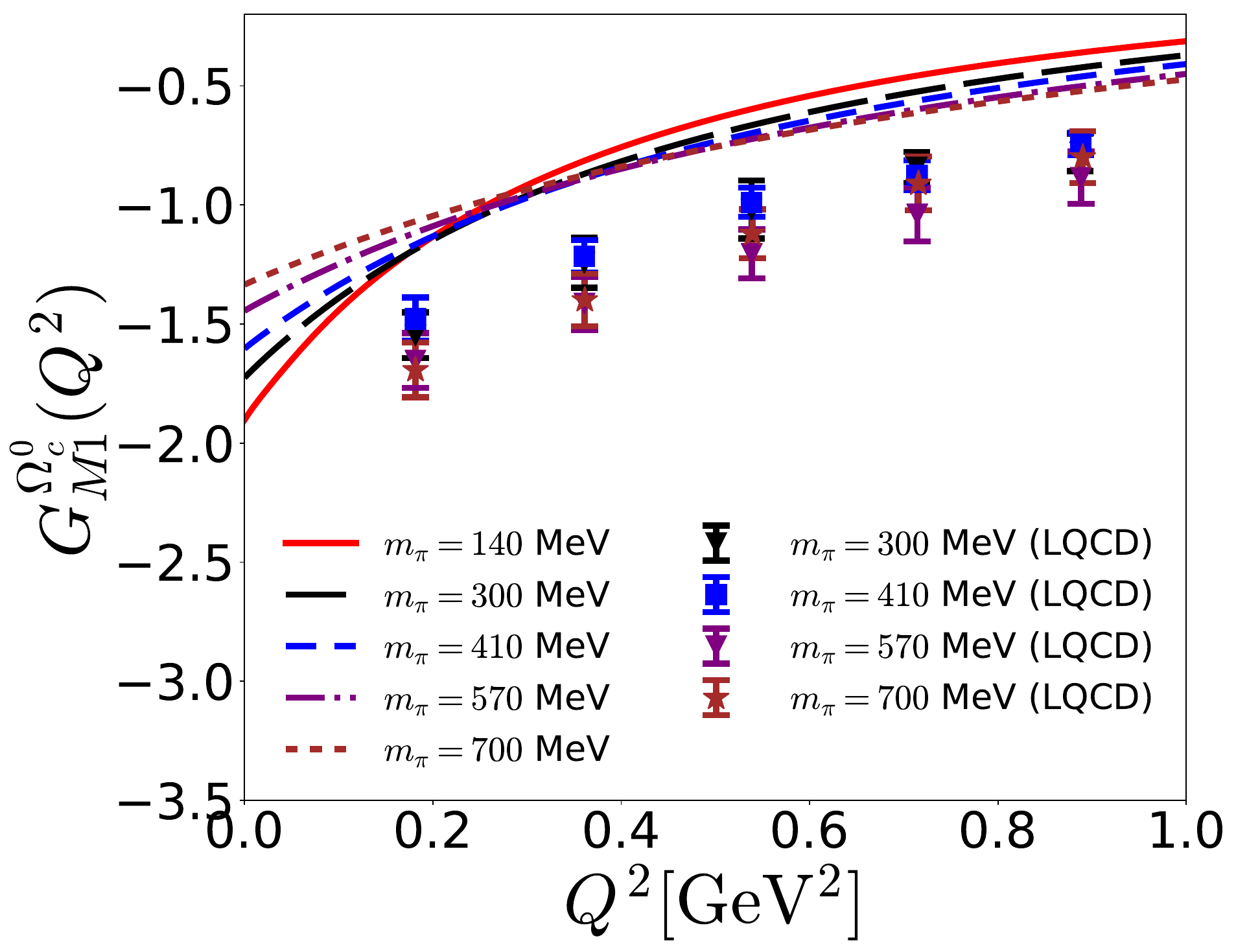}
  \caption{Results for the magnetic form factors of the singly heavy
    baryons.} 
  \label{fig:5}
\end{figure}
While there are no experimental data on the EM form factors of the
singly heavy baryons, the results from lattic QCD
exist~\cite{Can:2013tna, Bahtiyar:2016dom}. However, Can et
al.~\cite{Can:2013tna} employed the pion mass deviated from the
physical one, of which the value lies in the range of $300\le m_\pi
\le 700$ MeV. Thus, to compare the numerical results with those from
lattice data, we need to compute the EM form factors with $m_\pi$
varied. The results for the electric and magnetic form factors of the
singly heavy baryons with spin 1/2 are given in Figs.~\ref{fig:4}
and~\ref{fig:5}, respectively.  

When we increase the values of the pion mass, the results of the
electric form factors fall off more slowly than those with the
physical pion mass. Note that this is a well-known behavior of the
lattice results. On the other hand, as the pion mass becomes larger,
the electric form factors of the neutral heavy baryons increase
more slowly. If we increase the pion mass larger than 140
MeV, the Yukawa tail of the pion mean field is suppressed stronger,
which indicates that the size of the baryon gets more 
compact than the physical one. The dependence of the electric form
factors with unphysical pion masses reflect this fact.  
The numerical results for the $\Sigma_c^{++}$ electric form
factor are in good agreement with the lattice data. Those for 
$\Sigma_c^0$ and $\Omega_c^0$ get closer to the data with the pion mass
increased.  

To compare the $Q^2$ dependence of the magnetic form factors, we have
normalized the magnitudes of the magnetic form factors at $Q^2=0$ to
be the same as the lattice ones. Can et al.~\cite{Can:2013tna} carried
out the chiral extrapolation to the physical mass of the pion and
obtained the values of the quadratic fitting: $4.12$ for
$\Sigma_c^{++}$, $3.80$ for $\Sigma_c^0$, and $2.71$ for $\Omega_c$. 
We use them for normalization. The current results on the $Q^2$
dependence of the $M1$ form factors are generally in agreement with
the lattice data. We also observe that the lattice results lessen more
slowly, compared to the present ones. 

\begin{figure}[htp]
  \centering
  \includegraphics[scale=0.35]{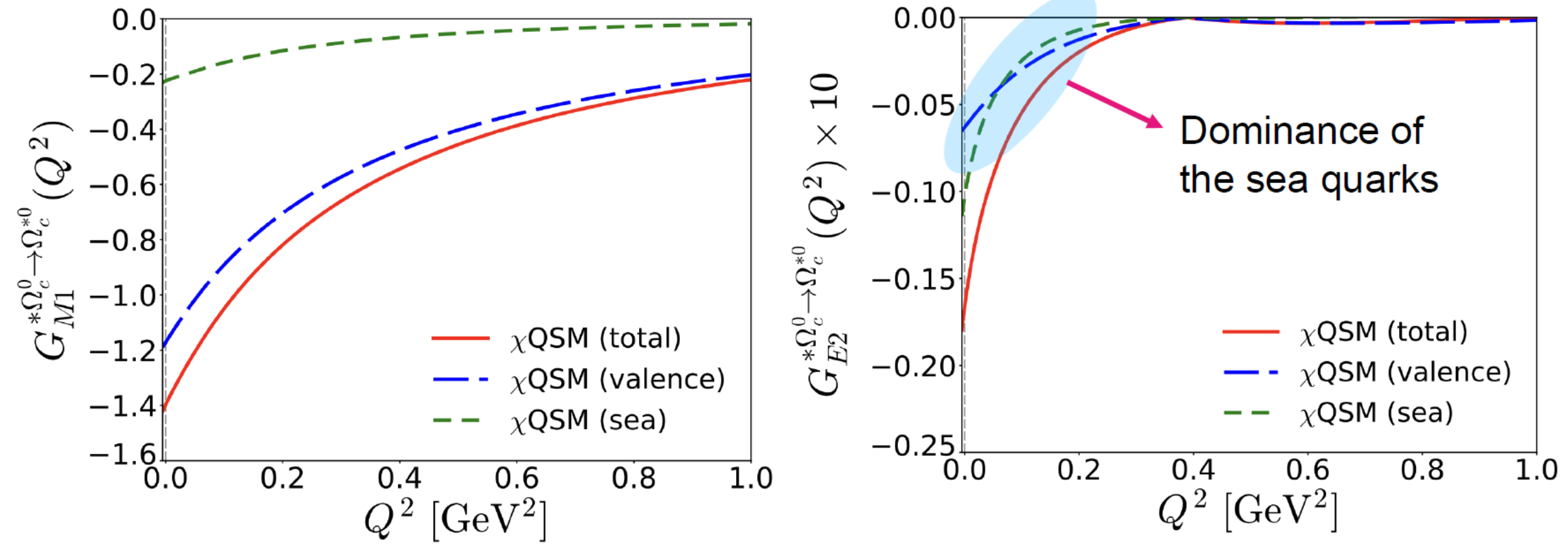}
  \caption{Results for the radiative $M1$ and $E2$ transition form
    factors for $\Omega_c\gamma \to \Omega_c^*$.} 
  \label{fig:6}
\end{figure}
We also computed the radiative transitions for the singly heavy
baryons with spin 3/2~\cite{Kim:2020uqo, Kim:2021xpp}. We will only
show the $\Omega_c^*$ transitions in this talk. As shown in
Fig.~\ref{fig:6}, the sea-quark contribution is dominant over that of
the valence quarks. As is well known from the $N\gamma \to \Delta$,
the electric quarupole form factor measures how the $\Delta$ is
deformed from a spherical shape. This implies that the outer part of
the $\Delta$ isobar comes into essential play to explain the $E2$
transition. This is a typical feature of the spin-3/2 baryon. On the
other hand, the sea-quark effects are marginal in the case of the $M1$
form factor. Though the lattice data on the EM transition form factors
of the singly heavy baryons exist~\cite{Bahtiyar:2016dom}, they suffer
from large uncertainty. 

\begin{table}[htp]
\setlength{\tabcolsep}{5pt} 
\global\long\def\arraystretch{1.2}%
 \caption{Results for the radiative decay widths of $B\gamma\to B^{*}$.}
   {\footnotesize
\begin{tabular}{cccccccc}
  \hline\hline
$\Gamma(B_{c}\gamma\to B_{c}^{*})$ 
  & $\chi$QSM
& $\chi$SM~\cite{Yang:2019tst} 
& LQCD~\citep{Bahtiyar:2015sga} 
& Bag~\citep{Bernotas:2013eia} 
& $\chi$PT~\citep{Wang:2018cre} 
& QCDSR~\citep{Aliev:2009jt,Aliev:2014bma} 
& QM~\citep{Ivanov:1999bk}
\tabularnewline
\hline 
$\Lambda_{c}^{+}\gamma\to\Sigma_{c}^{*+}$ 
& 69.76 
& $191.13\pm15.15$ 
& -- 
& 126 
& 161.8 
& 130(45) 
& 151(4)
\tabularnewline
$\Xi_{c}^{+}\gamma\to\Xi_{c}^{*+}$ 
& 31.97 
& $55.77\pm5.22$ 
& -- 
& 44.3 
& 21.6 
& 52(25) 
& 54(3)
\tabularnewline
$\Xi_{c}^{0}\gamma\to\Xi_{c}^{*0}$ 
& 0.08 
& $1.61\pm0.42$ 
& -- 
& 0.908 
& 1.84 
& 0.66(32) 
& 0.68(4)
\tabularnewline
\hline 
$\Sigma_{c}^{++}\gamma\to\Sigma_{c}^{*++}$ 
& 1.08 
& $2.41\pm0.22$ 
& -- 
& 0.826 
& 1.20 
& 2.65(1.20) 
& --
\tabularnewline
$\Sigma_{c}^{+}\gamma\to\Sigma_{c}^{*+}$ 
& 0.06 
& $0.11\pm0.02$ 
& -- 
& 0.004 
& 0.04 
& 0.40(16) 
& 0.140(4)
\tabularnewline
$\Sigma_{c}^{0}\gamma\to\Sigma_{c}^{*0}$ 
& 0.30 
& $0.80\pm0.06$ 
& -- 
& 1.08 
& 0.49 
& 0.08(3) 
& --
\tabularnewline
$\Xi_{c}^{\prime+}\gamma\rightarrow\Xi_{c}^{\ast+}$ 
& 0.09 
& $0.21\pm0.02$ 
& -- 
& 0.011 
& 0.07 
& 0.274 
& --
\tabularnewline
$\Xi_{c}^{\prime0}\gamma\rightarrow\Xi_{c}^{\ast0}$ 
& 0.34 
& $0.64\pm0.05$ 
& -- 
& 1.03 
& 0.42 
& 2.142 
& --
\tabularnewline
$\Omega_{c}^{0}\gamma\to\Omega_{c}^{*0}$ 
& 0.34 
& $0.49\pm0.08$ 
& 0.074 
& 1.07 
& 0.32 
& 0.932 
  & --

\tabularnewline
\hline \hline
\end{tabular}
}
\label{tab:1}
\end{table}
Table~\ref{tab:1} lists the present results in
the second column in comparison with those from other theoretical
works. Note that the results from Ref.~\cite{Yang:2019tst} is based on
the same theoretical framework, but all the dynamical parameters were
fixed in a \emph{model-independent} way by using the experimental data
on magnetic moments of hyperons~\cite{Yang:2015era}.

\section{Axial-vector structure of singly heavy baryons}
The strong decay rates for the singly heavy baryons can be derived by
considering the following collective operator~\cite{Kim:2017khv}:
\begin{align}
\mathcal{O}_{\varphi}=\frac{3}{M_{1}+M_{2}}
\sum_{i=1,2,3} \left[  G_{0}D_{\varphi\,i}^{(8)}-G_{1}\,d_{ibc}%
D_{\varphi\,b}^{(8)}\hat{S}_{c}-G_{2}\frac{1}{\sqrt{3}}D_{\varphi\,8}%
^{(8)}\hat{S}_{i}\right]  p_{i} ,  
\end{align}
where $p_i$ denotes the momentum of the outgoing meson of mass $m$.
Three coupling constants $G_i$ can be related to the dynamical
parameters $a_i$ for the axial-vector operator~\cite{Praszalowicz:1998jm,
  Yang:2015era} via the Goldberger-Treiman relation 
\begin{align}
\left\{  G_{0},G_{1},G_{2}\right\}  =\frac{M_{1}+M_{2}}{2f_{\varphi}}\frac
{1}{3}\left\{  -a_{1},a_{2},a_{3}\right\},  
\end{align}
where $F_\varphi$ stands for the meson decay constant ($f_\pi=93$ MeV or
$f_K=112$ MeV). The numerical values of $a_i$ can be determined by
using the experimental data on hyperon semileptonic
decays~\cite{Yang:2015era}:  
\begin{align}
a_{1}=-3.509\pm0.011,\;a_{2}=3.437\pm0.028,\;a_{3}=0.604\pm0.030.  
\label{eq:20}
\end{align}

The strong decay rate is expressed by 
\begin{align}
\Gamma_{B_1 \to B_2+\varphi} &= \frac1{2\pi} \overline{\langle
       B_2|\mathcal{O}_{\varphi}  |B_1\rangle^2} \frac{M_2}{M_1} p,
\end{align}
where $\overline{\langle \cdots\rangle}$ denotes the integration over
phase space. Thus, we obtain the results for the strong decay
rates. Tables~\ref{tab:2} and~\ref{tab:3} list the results for the
strong decay rates of the singly charmed and bottom baryons,
respectively, which are in remarkable agreement with the
experimental data~\cite{ParticleDataGroup:2024cfk}. We want to
emphasize that all the dynamical parameters have already been fixed in
the light baryon sector. 

\begin{table}[h]
\caption{Decay widths for the charm baryon sextet in MeV.}
\centering{}
 {\footnotesize
\begin{tabular}
[c]{crc}\hline\hline
 \text{Decay} & $\chi$QSM &
    \text{Exp.}\cite{ParticleDataGroup:2024cfk} \\\hline 
 $\Sigma_{c}^{++}(\boldsymbol{6}_{1},1/2)\rightarrow\Lambda_{c}%
^{+}(\overline{\boldsymbol{3}}_{0},1/2)+\pi^{+}$ & $1.93 $ & $1.89_{-0.18}%
^{+0.09}$\\
 $\Sigma_{c}^{+}(\boldsymbol{6}_{1},1/2)\rightarrow\Lambda_{c}%
^{+}(\overline{\boldsymbol{3}}_{0},1/2)+\pi^{0} $ & $2.24 $ & $<4.6$\\
 $\Sigma_{c}^{0}(\boldsymbol{6}_{1},1/2)\rightarrow\Lambda_{c}%
^{+}(\overline{\boldsymbol{3}}_{0},1/2)+\pi^{-} $ & $1.90 $ & $1.83_{-0.19}%
^{+0.11}$\\ 
 $\Sigma_{c}^{++}(\boldsymbol{6}_{1},3/2)\rightarrow\Lambda_{c}%
^{+}(\overline{\boldsymbol{3}}_{0},1/2)+\pi^{+} $ & $14.47 $ & $14.78_{-0.19}%
^{+0.30}$\\
 $\Sigma_{c}^{+}(\boldsymbol{6}_{1},3/2)\rightarrow\Lambda_{c}%
^{+}(\overline{\boldsymbol{3}}_{0},1/2)+\pi^{0} $ & $15.02  $ & $<17$\\
 $\Sigma_{c}^{0}(\boldsymbol{6}_{1},3/2)\rightarrow\Lambda_{c}%
^{+}(\overline{\boldsymbol{3}}_{0},1/2)+\pi^{-} $ & $14.49  $ & $15.3_{-0.5}%
^{+0.4}$\\ 
 $\Xi_{c}^{+}(\boldsymbol{6}_{1},3/2)\rightarrow\Xi_{c}(\overline
{\boldsymbol{3}}_{0},1/2)+\pi$ & $2.35  $ & $2.14\pm0.19$\\
 $\Xi_{c}^{0}(\boldsymbol{6}_{1},3/2)\rightarrow\Xi_{c}(\overline
{\boldsymbol{3}}_{0},1/2)+\pi$ & $2.53 $ & $2.35\pm0.22$\\\hline \hline
\end{tabular}
}%
\label{tab:2}%
\end{table}

\begin{table}[h]
\caption{Decay widths for the bottom baryon sextet in MeV.}
\centering{}
{\footnotesize 
\begin{tabular}
[c]{crc}\hline \hline
 \text{Decay} & $\chi$QSM  & \text{Exp.}
 \cite{ParticleDataGroup:2024cfk} \\\hline 
 $\Sigma_{b}^{+}(\boldsymbol{6}_{1},1/2)\rightarrow\Lambda_{b}%
^{0}(\overline{\boldsymbol{3}}_{0},1/2)+\pi^{+} $ & $6.12 $ & $9.7_{-3.0}%
^{+4.0}$\\
 $\Sigma_{b}^{-}(\boldsymbol{6}_{1},1/2)\rightarrow\Lambda_{b}%
^{0}(\overline{\boldsymbol{3}}_{0},1/2)+\pi^{-} $ & $6.12 $ & $4.9_{-2.4}%
^{+3.3}$\\ 
 $\Xi_{b}^{^{\prime}}(\boldsymbol{6}_{1},1/2)\rightarrow\Xi_{c}%
(\overline{\boldsymbol{3}}_{0},1/2)+\pi$ & $0.07 $ & $<0.08$\\ 
 $\Sigma_{b}^{+}(\boldsymbol{6}_{1},3/2)\rightarrow\Lambda_{b}%
^{0}(\overline{\boldsymbol{3}}_{0},1/2)+\pi^{+} $ & $10.96 $ & $11.5\pm2.8$\\
 $\Sigma_{b}^{-}(\boldsymbol{6}_{1},3/2)\rightarrow\Lambda_{c}%
^{0}(\overline{\boldsymbol{3}}_{0},1/2)+\pi^{-} $ & $11.77 $ & $7.5\pm
2.3$\\ 
 $\Xi_{b}^{0}(\boldsymbol{6}_{1},3/2)\rightarrow\Xi_{b}(\overline
{\boldsymbol{3}}_{0},1/2)+\pi$ & $0.80 $ & $0.90\pm0.18$\\
 $\Xi_{b}^{-}(\boldsymbol{6}_{1},3/2)\rightarrow\Xi_{b}(\overline
{\boldsymbol{3}}_{0},1/2)+\pi$ & $1.28 $ & $1.65\pm0.33$\\\hline \hline
\end{tabular}
}
\label{tab:3}%
\end{table}

Using the parameters given in Eq.~\eqref{eq:20}, we can also study the
quark spin content of the baryon sextet~\cite{Suh:2022atr}. Since the
antitriplet baryons consist of the $J=0$ pion mean field and a heavy
quark, no light-quark contributions exists for them. In this talk, we
only present the results for the quark spin content of the baryon
sextet with both spin $J'=1/2$ and $J'=3/2$. Tables~\ref{tab:4}
and~\ref{tab:5} list the corresponding results, respectively. Compared
with the lattice data~\cite{Alexandrou:2016xok}, the results are in
agreement with them. For detailed formalism and discussion, we refer
to Ref.~\cite{Suh:2022atr}. 

\begin{table}[htp]
\centering
\caption{Quark spin content of the baryon sextet with $J'=1/2$ } 
\label{tab:4}
 {\footnotesize
 \begin{tabular}{ c | c c c c c } 
  \hline 
    \hline 
 $J'_{3}=1/2$ & $g^{(0)}_{A}$  & $\Delta{u}$ & $\Delta{d}$ & $\Delta{s}$ & $\Delta{c}$ \\
  \hline 
$\Sigma^{++}_{c}$ & $0.566$ & $0.991$ & $-0.064$  &  $-0.028$  & $-0.333$ \\    
$\Sigma_{c}$~\cite{Alexandrou:2016xok}&$ 0.4094 \pm 0.0199 $ 
& $0.7055 \pm 0.0191$ & $-$  &  $-$  & $-0.2970 \pm 0.0113$ \\  
$\Xi^{\prime +}_{c}$&$0.531$ & $0.505$ & $-0.087$  &  $0.447$  & $-0.333$ \\  
$\Xi^{\prime}_{c}$~\cite{Alexandrou:2016xok}&$ 0.4872 \pm 0.0127 $ 
& $0.3433 \pm 0.0085$ & $-$  &  $ 0.4539 \pm 0.0055 $  & $-0.3133 \pm 0.0069$ \\
$\Omega^{0}_{c}$& $0.497$ & $-0.069$ & $-0.069$  &  $0.968$  & $-0.333$ \\  
$\Omega^{0}_{c}$~\cite{Alexandrou:2016xok}&$ 0.5428 \pm 0.0118 $ 
& $-$ & $-$  &  $ 0.8554 \pm 0.0117 $  & $-0.3125 \pm 0.0054$ \\
 \hline 
 \hline
\end{tabular}
 }
\end{table}

\begin{table}[htp]
\centering
\caption{Quark spin content of the baryon sextet with $J'=3/2$}  
\label{tab:5}
 {\footnotesize 
 \begin{tabular}{ c | c c c c c } 
  \hline 
    \hline 
$J'_{3}=3/2$  & $g^{(0)}_{A}$  & $\Delta{u}$ & $\Delta{d}$ &
 $\Delta{s}$ & $\Delta{c}$ \\ 
  \hline 
$\Sigma^{* ++}_{c}$ & $2.349$ 
& $1.487$ & $-0.096$  &  $-0.042$  & $1.000$ \\  
 $\Sigma^{*}_{c}$~\cite{Alexandrou:2016xok}&$ 2.0004 \pm 0.0346 $ 
& $1.0899 \pm 0.0308$ & $-$  &  $-$  & $0.9043 \pm 0.0090$ \\
$\Xi^{* +}_{c}$&$2.297$ & $0.889$  &  $-0.131$  &  $0.670$  & $1.000$ \\ 
$\Xi^{*}_{c}$~\cite{Alexandrou:2016xok}&$ 2.1192 \pm 0.0254 $ & $0.5466 \pm 0.0150$  & $-$  &  $0.6587 \pm 0.0104$  & $0.9103 \pm 0.0075$ \\ 
$\Omega^{* 0}_{c}$ &$2.245$ & $-0.104$ & $-0.104$ &  $1.452$  & $1.000$ \\  
$\Omega^{* 0}_{c}$~\cite{Alexandrou:2016xok} &$ 2.1961 \pm 0.0261 $ & $-$ & $-$  &  $1.2904 \pm 0.0204$  & $0.9026 \pm 0.0090$ \\
 \hline 
 \hline
\end{tabular}
 }
\end{table}
\section{Summary and conclusions}
In the current talk, we first discussed a pion mean-field approach or
the chiral quark-soliton model, which describe both the low-lying
light baryons and singly heavy baryons on an equal footing. We first
showed the results for the electromagnetic form factors of the singly
heavy baryons and compared them with the lattice data. Once we
introduce large unphysical values of the pion mass, which correspond
to those used in lattice QCD, the results are in agreement
with the lattice data. We also presented the results for the radiative
and strong decay widths for the singly heavy baryons. Finally, we
showed the results for the quark spin content of the baryon sextet
with both spin $J'=1/2$ and $J'=3/2$, which are in qualitative
agreement with the lattice data. 

We considered the infinitely large mass of the heavy quark in
the current talk. we want to mention that to introduce the $1/m_Q$
corrections one has to deal with gluonic degrees of freedom. Very
recently, we constructed a new effective theory for the baryon from
the instanton vacuum~\cite{Choi:2025xha}, within which we can derive
an effective operator for a gluonic operator. We will soon compute
various observables for the singly heavy baryons with the $1/m_Q$
corrections. 

\acknowledgments
The current work was a summary of a series of the works done together
with Y.-S. Jun, J.~Y. Kim, M. Oka, J.~M. Suh, and G.~S. Yang. I want
to express my gratitude to them for fruitful collaborations.
The present work was supported by the Basic Science Research Program
through the National Research Foundation of Korea funded by the Korean
government (Ministry of Education, Science and Technology, MEST),
Grant-No. 2021R1A2C2093368 and 2018R1A5A1025563.

\end{document}